\documentclass[fleqn,usenatbib]{mnras}

\usepackage[T1]{fontenc}

\DeclareRobustCommand{\VAN}[3]{#2}
\let\VANthebibliography\thebibliography
\def\thebibliography{\DeclareRobustCommand{\VAN}[3]{##3}\VANthebibliography}

\usepackage{graphicx}	
\usepackage{amsmath}	
\usepackage{amssymb}	
\usepackage{mathpazo}

\def\la{\mathrel{\hbox{\rlap{\hbox{\lower4pt\hbox{$\sim$}}}\hbox{$<$}}}}
\def\ga{\mathrel{\hbox{\rlap{\hbox{\lower4pt\hbox{$\sim$}}}\hbox{$>$}}}}

\title[MeerKAT detection of LAMOST J0240]{Radio and optical observations of the possible AE Aqr twin, LAMOST J024048.51+195226.9}

\author[M.L. Pretorius et al.]{
M. L. Pretorius,$^{1}$\thanks{E-mail: retha@saao.ac.za (MLP)}
D. M. Hewitt,$^{1,2,3}$
P. A. Woudt,$^{2}$
R. P. Fender,$^{4,2}$
I. Heywood,$^{4,5,6}$
C. Knigge,$^{7}$
\newauthor J. C. A. Miller-Jones,$^8$
D. A. H. Buckley,$^{1}$
H. L. Worters,$^{1}$
S. B. Potter,$^{1,9}$
D. R. A. Williams$^{10}$
\\
$^{1}$South African Astronomical Observatory, PO Box 9, Observatory 7935, South Africa\\
$^{2}$Department of Astronomy, University of Cape Town, Private Bag X3, Rondebosch 7701, South Africa\\
$^{3}$Anton Pannekoek Institute for Astronomy, University of Amsterdam, Science Park 904, 1098 XH Amsterdam, The Netherlands\\
$^{4}$Department of Physics, Astrophysics, University of Oxford, Denys Wilkinson Building, Keble Road, Oxford OX1 3RH, UK\\
$^{5}$Department of Physics and Electronics, Rhodes University, PO Box 94, Makhanda 6140, South Africa\\
$^{6}$South African Radio Astronomy Observatory, 2 Fir Street, Black River Park, Observatory, Cape Town 7925, South Africa\\
$^{7}$School of Physics and Astronomy, University of Southampton, Highfield, Southampton, SO17 1BJ, UK\\
$^8$International Centre for Radio Astronomy Research -- Curtin University, GPO Box U1987, Perth, WA 6845, Australia\\
$^9$Department of Physics, University of Johannesburg, PO Box 524, Auckland Park 2006, South Africa\\
$^{10}$Jodrell Bank Centre for Astrophysics, School of Physics and Astronomy, The University of Manchester, Manchester, M13 9PL, UK
}
\date{Accepted XXX. Received YYY; in original form ZZZ}

\pubyear{2021}

\begin{document}
\label{firstpage}
\pagerange{\pageref{firstpage}--\pageref{lastpage}}
\maketitle

\begin{abstract}
  \citet{Thorstensen2020} recently argued that the cataclysmic variable (CV) LAMOST J024048.51+195226.9 may be a twin to the unique magnetic propeller system AE Aqr. If this is the case, two predictions are that it should display a short period white dwarf spin modulation, and that it should be a bright radio source. We obtained follow-up optical and radio observations of this CV, in order to see if this holds true. Our optical high-speed photometry does not reveal a white dwarf spin signal, but lacks the sensitivity to detect a modulation similar to the 33-s spin signal seen in AE Aqr. We detect the source in the radio, and measure a radio luminosity similar to that of AE Aqr and close to the highest so far reported for a CV. We also find good evidence for radio variability on a time scale of tens of minutes. Optical polarimetric observations produce no detection of linear or circular polarization. While we are not able to provide compelling evidence, our observations are all consistent with this object being a propeller system.
\end{abstract}




\section{Introduction}
CVs are semi-detached binaries, in which a white dwarf (WD) accretes from a low-mass donor star (see e.g.\ \citealt{warner}). In magnetic CVs, the WD has a magnetic field sufficiently strong to alter the accretion flow significantly. These systems are divided into two subtypes, namely polars and intermediate polars (IPs).

Polars are characterized by strong circular and linear polarization, modulated at the orbital period ($P_{orb}$), indicating WD rotation that is synchronized (or very close to synchronized) with the binary orbit. IPs have highly coherent pulsations, at periods $<P_{orb}$, in their X-ray and/or optical light curves, interpreted as the spin modulation of a WD rotating at a period much shorter than the orbit. 

\citet{Patterson1979} discovered very stable optical pulsations at a period of 33.07~s in the nearby CV AE Aqr, leading him to suggest that it is an ``oblique rotator'' (a system in which the WD spin and magnetic axes are misaligned, and the accretion flow is magnetically channeled from the inner edge of a truncated accretion disc, onto the WD), the model now used to describe other IPs. \cite{deJager1994} found that the WD in AE Aqr is spinning down at a high rate ($5.6\times 10^{-14}\,{\rm s/s}$). AE Aqr was also one of the first CVs to be detected in the radio \citep{BookbinderLamb1987}, and at  $\sim 5 \times 10^{16}\,{\rm erg\,s^{-1}\,Hz^{-1}}$ is amongst the most radio luminous CVs \citep[e.g.][]{Bastian1988, Barrett2017, CoppejansKnigge2020}. The radio emission is modelled as a superposition of synchrotron sources \citep{Bastian1988}.

Besides the 33-s signal, the observed behaviour for which AE Aqr is best known is its spectacular flaring. Many observers have reported large-amplitude aperiodic flaring in optical, X-ray, and radio light curves (on time scales of minutes to tens of minutes), and also in the optical emission lines of this system \citep[e.g.][]{Patterson1979, Bastian1988, Skidmore2003, Welsh1998, ChoiDotani1999, ChoiDotani2006}.

The rapid spin-down, with spin-down power greatly exceeding the radiated power, the lack of evidence for a disc, and the spectacular flares seen in AE Aqr are explained by a ``magnetic propeller'' where the rapidly rotating, strongly magnetic WD expels most of the infalling gas, preventing it from accreting \citep{EracleousHorne1996, WynnKingHorne1997}. AE Aqr remains the only CV that is modelled as a magnetic propeller, and up until recently, had the shortest known WD spin period\footnote{In the last few months, very short WD spin periods were identified in two CVs, CTCVJ2056-3014 and V1460 Her (29.6 and 38.6~s, respectively). These systems however both have at least partial discs and are accretion powered; i.e.\ apart from harbouring rapidly spinning WDs, they are similar to normal IPs, rather than to AE Aqr \citep{LopesdeOliveira2020, Ashley2020}.}.

The recently discovered LAMOST J024048.51+195226.9, also catalogued as CRTS J024048.5+195227, (hereafter J0240) may be another example of a magnetic propeller. \cite{Thorstensen2020} points out several optical properties reminiscent of AE Aqr, including large-amplitude flares on time scales down to $\sim$1 minute, weak or absent He\,{\scriptsize II}\,$\lambda$4686 emission (usually very strong in magnetic CVs, but not in AE Aqr), and emission lines that show irregular variations, with radial velocities that do not seem to trace the orbit. The system has an orbital period of 7.3 hours \citep{Drake2014, Thorstensen2020, LittlefieldGarnavich2020}.

If J0240 really is another propeller system, an observational signature would be a rapid WD spin modulation in optical and/or X-ray light curves (the light curves of \cite{Thorstensen2020} had only 23 and 30~s cadence). One would also expect it to show bright radio emission, by analogy with AE Aqr. Here we present the first radio detection of J0240, as well as optical high-speed photometry and photo-polarimetry. Our observations are described in Section~\ref{sec:observations}. The results are presented and discussed in  Section~\ref{sec:results}. Finally, Section~\ref{sec:summary} summarizes our work.

\section{Observations}
\label{sec:observations}
We obtained radio and optical data of J0240 in the second half of 2020, using the MeerKAT radio interferometer, the South African Astronomical Observatory (SAAO) 1-m optical telescope, Lesedi, and the SAAO 74-inch telescope. Table~\ref{tab:obslog} gives a log of the observations.

\subsection{MeerKAT radio interferometer}
A roughly 2 hour observation was taken with MeerKAT on 12 Aug 2020, as part of the ThunderKAT survey (The Hunt for Dynamic and Explosive Radio Transients using MeerKAT; \citealt{Fender2017}). The observation was done in the L-band (centered at 1284 MHz, with a bandwidth of 856 MHz covered by 4096 channels), using 59 of the 64 antennas. It started and ended with scans of the primary calibrator (J0408-6545), and alternated between the secondary calibrator (J0238+1636) and target for the rest of the observation. Visibilities were recorded every 2 seconds, and the 6 target scans were each approximately 15 minutes in length.

The data were calibrated and imaged using the OxKAT\footnote{Available from https://github.com/IanHeywood/oxkat/} pipeline \citep{oxcat}. This implements standard \textsc{casa} \citep[e.g.][]{McMullin2007} routines, as well as the SARAO tricolour flagger\footnote{https://github.com/ska-sa/tricolour/}, and the imaging packages DDFacet \citep{ddfacet} and WSCLEAN \citep{wsclean}. We found that second generation calibration (direction independent self-calibration) does not significantly improve the image quality in the region near the target, and therefore report results obtained from applying only first generation calibration. Noise and flux density measurements were performed with PyBDSF \citep{blobs}.

\subsection{Optical photometry with the Lesedi 1-m telescope}
We observed J0240 on three consecutive nights in September 2020 with the new 1-m optical telescope, Lesedi, at the SAAO site in Sutherland. We performed high-speed photometry in white light (i.e.\ filterless photometry), using the imager SHOC \citep{shoc1, shoc2}. The integration time was 5 s on all nights, and since this is a frame-transfer CCD, there was no dead-time between integrations. Only quite short ($<3$ hours) observations were possible at the northern declination of J0240, and this early in the season.

We used the TEA-Phot\footnote{https://bitbucket.org/DominicBowman/tea-phot/src/master/} code \citep{BowmanHoldsworth2019} to extract the photometry. Since no filter was used, it is not possible to precisely place these data on a standard photometric system. However, by comparing the instrumental magnitudes of several stars in the field with catalogued $B$- and $R$-band measurements, we are able to perform a very rough calibration (to within $0.1$~mag), and verify that J0240 was at about the same brightness as during the observations of \citet{Thorstensen2020}, just above 17th magnitude for most of the time\footnote{Long-term CRTS, ASAS-SN, and ATLAS photometry also show no evidence of outbursts or low states in J0240.}.

\subsection{Optical all-Stokes Polarimetry with the 74-inch telescope}

Photo-polarimetry of J0240 was obtained on a single night in November 2020, over a period of $\sim$1 hour, using the SAAO 74-inch telescope with the HIgh-speed Photo-POlarimeter (HIPPO; \citealt{hippo}). This instrument performs time-resolved, simultaneous all-Stokes observations. No filter was used, implying a very broad bandpass (roughly 3500 to 9000~\AA). The polarimetric data were reduced as described in \cite{hippo}.

\begin{table}
	\centering
	\caption{Log of the observations of J0240. The MeerKAT radio observations used J0408-6545 as primary calibrator and J0238+1636 as secondary calibrator. For the Lesedi photometry, the integration time (and time resolution) was 5~s on all nights.}
	\label{tab:obslog}
	\begin{tabular}{ll} 
		\hline
                Start Date and Time (UTC) & Total time on target (hours) \\
		\hline
		\multicolumn{2}{l}{\textbf{MeerKAT}} \\
		2020 Aug 12 00:18:49.7    & 1.501 \\[0.2cm]
                \multicolumn{2}{l}{\textbf{Lesedi + SHOC}} \\
                2020 Sep 8 00:19:46.0     & 2.499 \\
		2020 Sep 9 00:56:22.0     & 2.653 \\
		2020 Sep 10 00:07:45.0    & 2.012 \\[0.2cm]
                \multicolumn{2}{l}{\textbf{74-inch + HIPPO}} \\
                2020 Nov 12 22:27:41.0    & 1.117 \\
		\hline
	\end{tabular}
\end{table}

\section{Results and Discussion}
\label{sec:results} 

\subsection{The radio data}

\begin{figure}
  	\includegraphics[width=\columnwidth]{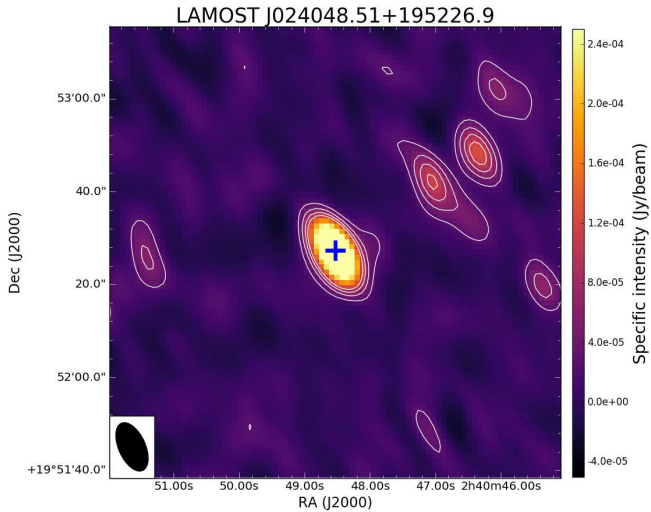}
        \caption{A $100'' \times 100''$ area of the MeerKAT radio map of the J0240 field, with north up and east to the left. The RMS noise in the vicinity of the point source at the center is $9.0\,\mu {\rm Jy\, beam^{-1}}$, and contours are at 3, 6, 9, and 12 $\times$ the RMS value. The blue cross marks the optical position of J0240, as measured by \emph{Gaia} (the size of this symbol does not denote anything). The inset in the bottom left-hand corner shows the beam shape and size ($11.42'' \times 6.11''$ with a  position angle of $29.2^\circ$).}
    \label{fig:radiomap}
\end{figure}

The radio map, with contours overlayed, of a small area at the center of the MeerKAT field is shown in Fig.~\ref{fig:radiomap}. We detect a bright radio point source coincident with the optical position of J0240, and assume it is the CV. The integrated radio flux density of J0240 is $0.60 \pm 0.02\,{\rm mJy}$, and the RMS noise measured nearby is $9.0\,\mu {\rm Jy\, beam^{-1}}$.

Splitting the 865 MHz bandwidth into 8 frequency sub-bands, we measure an in-band spectral index of $\alpha =-0.6 \pm 0.2$ (where $S_\nu \propto \nu^\alpha$). Note, however, that there are still bandpass calibration uncertainties for MeerKAT, and also that the source is variable (see below).
In the case of AE Aqr, the time averaged spectral index is positive \citep{Abada-Simon1993, Dubus2007}, but it varies rapidly, as may be expected from a changing optical depth in synchrotron emitting blobs \citep{Bastian1988}.

\begin{figure*}
	\includegraphics[width=160mm]{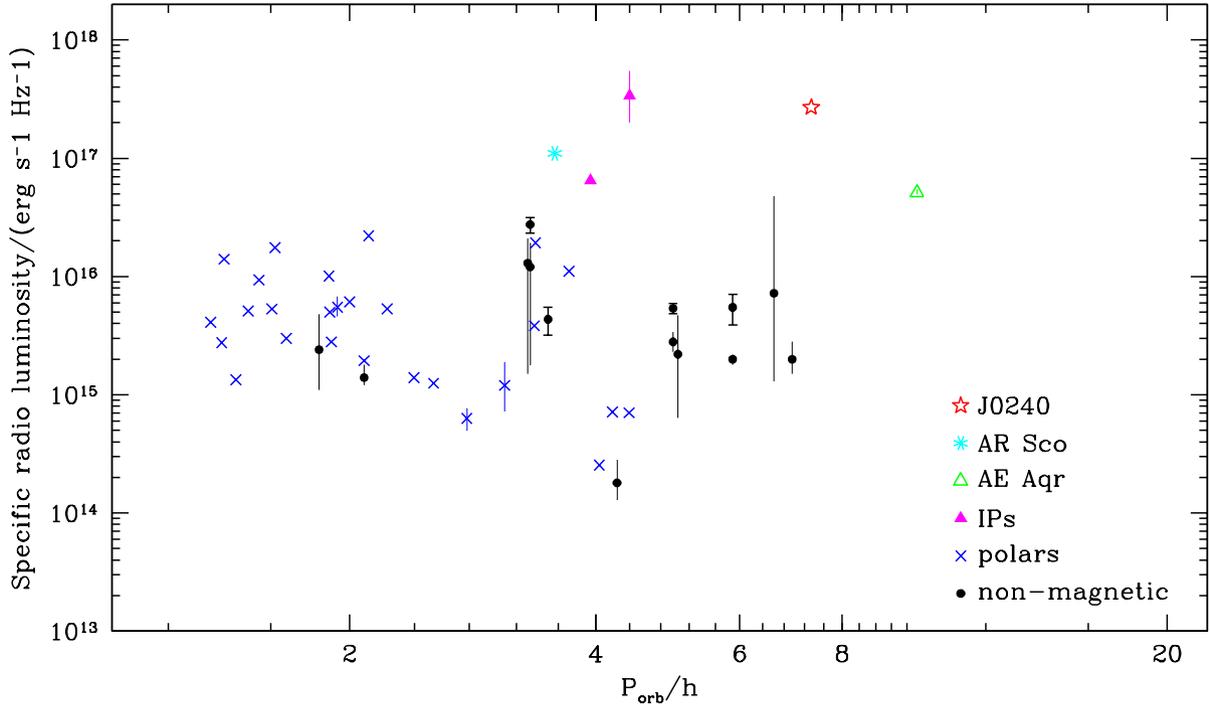}
	\caption{Radio luminosities as a function of orbital period for 42 CVs and the WD pulsar AR Sco. The CV classes are distinguished by different symbols. Some of these systems have been observed at a range of radio flux densities, which we indicate with the fine vertical bars. The error bars for J0240 are not shown, but are smaller than the symbol size. Points with error bars are the L-band observations from \citet{Hewitt2020}; other data are at higher frequencies, mainly C- and X-band, and are from \citet{Coppejans2016, Coppejans2015, Kording2008, MillerJones2011, Russell2016, Marsh2016, Marcote2017, Barrett2017, Barrett2020}. }
    \label{fig:radiolum}
\end{figure*}

At the \emph{Gaia} Early Data Release 3 \citep{gaia, gaiadr3} distance of $620 \pm 30\,{\rm pc}$, J0240 has a specific radio luminosity of $2.7 \pm 0.3 \times 10^{17}\,{\rm erg\,s^{-1}\,Hz^{-1}}$, placing it amongst the most radio-luminous CVs. Fig.~\ref{fig:radiolum} shows radio luminosity as a function of orbital period for CVs belonging to different classes (see \citealt{CoppejansKnigge2020} for an earlier version of this plot). Besides J0240, this includes all CVs with known distances and recent, sensitive radio observations between $\sim$1 and $12\,{\rm GHz}$. In addition, we include the WD pulsar, AR Sco \citep{Marsh2016, Marcote2017}. The highest point in this plot (belonging to the IP V1323 Her) is at slightly lower luminosity than shown in \citet{CoppejansKnigge2020}, because the distance was revised down (from 2240 to 1950~pc) in the \emph{Gaia} EDR3.

While J0240, AE Aqr, and the two normal IPs shown in Fig.~\ref{fig:radiolum} are all similarly radio-luminous, IPs are detected at radio wavelengths at a rate of $\la$10\%. \cite{Barrett2017} obtained sensitive VLA observations of more than 40 IPs and candidate IPs, including several well-studied systems at smaller distances than J0240, and (besides AE Aqr) detected only two. These two IPs, V1323 Her and Cas 1 (also known as RX J0153.3+7446) are both at large distances ($\ga 1.5\,{\rm kpc}$), and have only low $S/N$ detections in only a small subset of the VLA observations \citep{Barrett2017}.

\begin{figure}
  \includegraphics[width=\columnwidth]{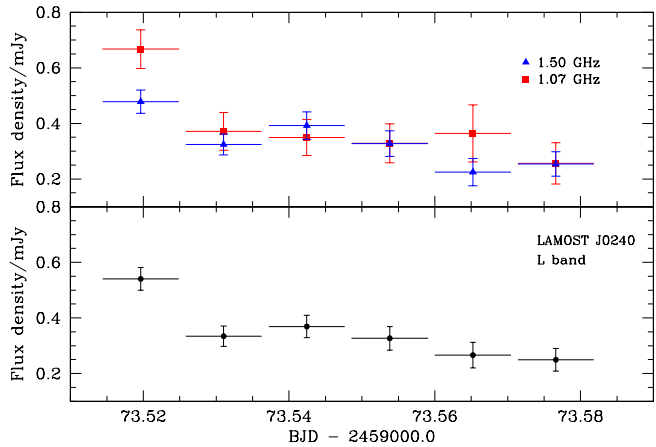}
    \caption{The radio light curve of J0240, constructed from the data shown in Fig.~\ref{fig:radiomap}. Each time bin is close to 15 minutes. The bottom panel shows the full L-band data, while in the top panel, this was split into 2 frequency sub-bands, centered on 1.07~GHz and 1.50~GHz.}
    \label{fig:radiolc}
\end{figure}

In order to look for radio variability, we split the observations into six 15 minute time bins, corresponding to the on-target scans, and imaged each of these epochs separately. Fig.~\ref{fig:radiolc} shows the resulting radio light curves. The bottom panel is the integrated radio flux density as a function of time, showing that J0240 varies by $>5 \sigma$, on a timescale of tens of minutes.
The top panel of Fig.~\ref{fig:radiolc} shows each time bin further split into two equal frequency bins.

If we assume that the radio emission is due to synchrotron radiation, and that we caught the end of a radio flare that evolved from optically thick to optically thin because of a decreasing optical depth to synchrotron self absorption, we can estimate a minimum energy following \cite{FenderBright2019}. Assuming further that the flare peaked at 0.8 mJy (at 1.5 GHz), then, for an integrated radio luminosity of $2.2 \times 10^{31}\,{\rm erg\,s^{-1}}$, the minimum energy is $1.5\times 10^{35}\,{\rm erg}$. This value corresponds to an emitting region size of $1.0 \times 10^{12}\,{\rm cm}$ and a plasma magnetic field of $0.53\,{\rm G}$. This energy is higher than the minimum energy found for a radio flare from SS Cyg by \citet{Fender2019}, but the uncertainty on our estimate is at least an order of magnitude.

\subsection{The optical high-speed photometry}

\begin{figure}
	\includegraphics[width=\columnwidth]{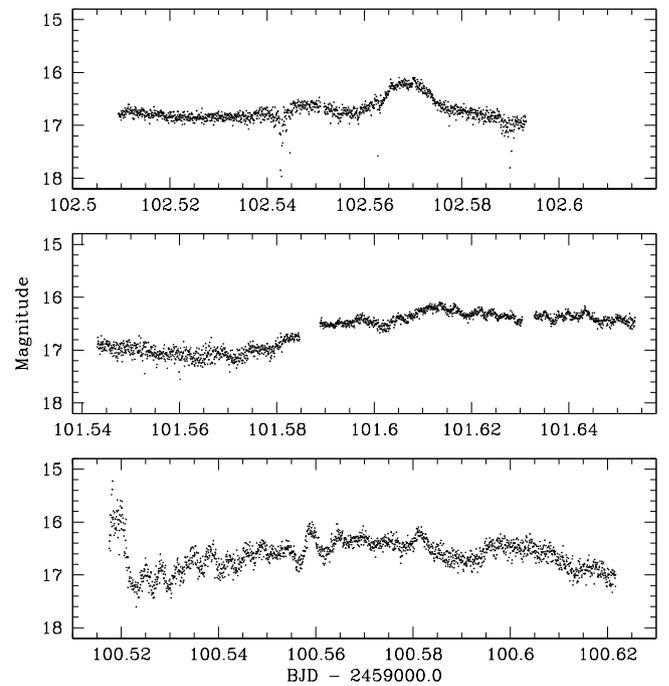}
    \caption{The unfiltered optical light curves of J0240, at 5-s time resolution, on 3 consecutive nights. Seeing conditions were especially poor on the third night (top panel). The very low points in that light curve are measured in frames with very bad image quality; these spurious points were removed before the Fourier transforms were calculated.}
    \label{fig:optlc}
\end{figure}

The optical light curves of J0240 are displayed in Fig.~\ref{fig:optlc}. These are all differential light curves, implying that colour differences between J0240 and comparison stars were ignored in correcting the photometry for atmospheric extinction. The system displays the rapid flickering typically seen in CVs (the observational signature of mass transfer), but these short runs show very little of the flaring reported by \cite{Thorstensen2020}. The photometry is also of fairly low quality, since it was obtained mainly in conditions of poor seeing.

Our Nyquist frequency of 0.1~Hz should be sufficiently high to allow a detection of the WD spin signal. A higher spin frequency would imply a WD above $1\,{\rm M}_\odot$ (see e.g.\ \citealt{Otoniel2020}). While this is of course possible, there is nothing to indicate that the WD in J0240 is more massive than the $\simeq 0.8\,{\rm M}_\odot$ more commonly measured for WDs in CVs (a mass that would imply a break-up spin period closer to 20~s).

\begin{figure}
	\includegraphics[width=\columnwidth]{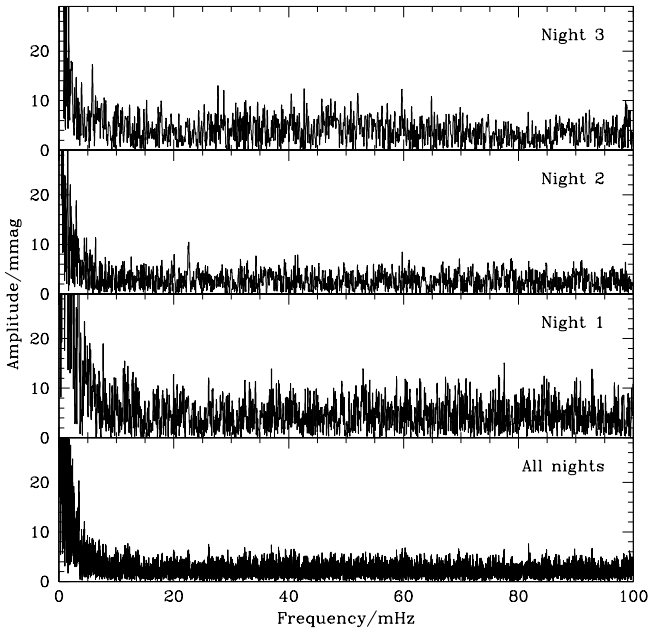}
    \caption{Fourier transforms of the optical light curves of J0240. In the top 3 panels, Fourier transforms of the light curves taken on each of the 3 nights are shown individually, while the Fourier transform in the bottom panel is of all data combined.}
    \label{fig:fts}
\end{figure}

\begin{figure}
	\includegraphics[width=\columnwidth]{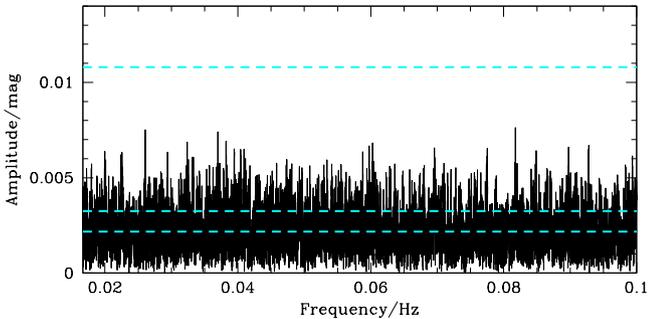}
        \caption{The Fourier transform of all 3 nights of photometry combined, showing only the frequency range corresponding to periods between 60 and 10~s. Dashed horizontal lines indicate amplitudes of 0.2, 0.3, and 1\%.}
    \label{fig:ftzoom}
\end{figure}

Fig.~\ref{fig:fts} shows discrete Fourier transforms of our optical light curves. We detect no coherent signals. It would, however, not be at all surprising if this system is in future found to have an optical modulation similar to the 33-s signal in AE Aqr. In his optical data of AE Aqr, \cite{Patterson1979} observed an amplitude exceeding 1\% (corresponding to $2.5 \log(101/100)\simeq 0.01\,{\rm mag}$) at times during flares, but the average amplitude was only 0.2 to 0.3\%. \cite{BeskrovnayaIkhsanovBruch1995} also reported no detection in photometry spanning 4 weeks, implying a conservative upper limit amplitude of $<0.005$ mag.
Fig.~\ref{fig:ftzoom} zooms in on the high frequency end of the Fourier transform of the combined J0240 light curve (this of course has lower white noise than the Fourier transforms of individual light curves, but actually also has less sensitivity to a signal that is present for only a short while). This shows that while there is no persistently present signal at the 1\% level in these data, a 0.2 to 0.3\% signal would be well below the noise level. Our photometry therefore cannot rule out a rapidly spinning WD giving rise to a modulation very similar to the one observed in AE Aqr.

For the interpretation of J0240 as a magnetic propeller, it remains key to demonstrate that this system contains a rapidly spinning WD. Longer optical light curves may show this. Alternatively, X-ray or $UV$ timing would be worth obtaining (J0240 has not yet been detected at X-ray or $UV$ frequencies). The 33-s signal of AE Aqr has an amplitude $>$20\% in X-rays \citep{Patterson1980}.

\subsection{The optical polarimetry}

There was no significant detection of polarization. Our measurement of time-averaged circular polarization is $0.10\% \pm 0.07\%$, consistent with a value of zero. For the time-averaged linear polarization, we find $1.3\% \pm 0.1\%$, consistent with an interstellar (or instrumental) origin.

Polarimetric observations of AE Aqr have yielded possible, marginal detections of circular polarization, but at levels of $\la 0.1\%$ (e.g.\ \citealt{Cropper1986}; \citealt{Beskrovnaya1996}; \citealt{Butters2009}).

\section{Summary}
\label{sec:summary}
We have presented radio L-band imaging, and optical high-speed photometry and photo-polarimetry of the recently discovered CV, J0240. Our aim was to examine the similarity of J0240 to the magnetic propeller system, AE Aqr. The high radio luminosity supports the suggestion that this system may be an AE Aqr-like object. However, the most important question, which we are unable to answer, is whether it contains a rapidly spinning WD.
The main results of this study are listed below.

\begin{enumerate}
\item J0240 is detected as a bright radio source. We measure a 1.284 GHz flux density of $0.60 \pm 0.02\,{\rm mJy}$, and an in-band spectral index of $-0.6 \pm 0.2$.
\item The radio luminosity of $2.7 \pm 0.3 \times 10^{17}\,{\rm erg\,s^{-1}\,Hz^{-1}}$ is the second highest yet reported for a CV.
\item The system varies in the radio on a time scale of 10s of minutes.
\item We fail to detect any coherent signal that can be attributed to a rapidly spinning WD in our 3 nights of optical high-speed photometric observations.
\item Although we are able to search for signals with periods down to 10~s, the sensitivity of our high-speed photometry to low amplitude signals is not sufficient to rule out a WD spin modulation similar to the 33-s signal seen in AE Aqr.
\item There was no detection of linear or circular polarization in the optical.
\end{enumerate}  

\section*{Acknowledgements}
MLP acknowledges financial support from the National Research Foundation (NRF) and the Newton Fund. PAW acknowledges financial support from the University of Cape Town and the NRF. We thank John Thorstensen for helpful discussions of the system that this work focusses on. Sara Motta kindly looked at the archived \emph{Swift} data of this object, to confirm that there is no X-ray detection. The MeerKAT telescope is operated by the South African Radio Astronomy Observatory (SARAO), which is a facility of the NRF, an agency of the Department of Science and Innovation. We thank the SARAO staff involved in obtaining the MeerKAT observations. We made use of the computing facilities of the Inter-University Institute for Data Intensive Astronomy (IDIA) in this research. 

\section*{Data Availability}
The data presented in this article are subject to the standard data access policies of the South African Radio Astronomy Observatory and the South African Astronomical Observatory.







\bsp	
\label{lastpage}
\end{document}